%% file: WCNC2011_KS_low_complexity_vf.tex
\documentclass[conference]{IEEEtran}
\input{pream_bm}

\usepackage{epsfig,latexsym,amssymb,amsmath,graphics}
\usepackage{graphicx}

\usepackage{amsfonts}
\usepackage{dsfont}

\begin{document}

%
\title{Low Complexity Kolmogorov-Smirnov Modulation Classification}

\author{\IEEEauthorblockN{Fanggang Wang$^{*\dag}$,\ Rongtao Xu$^\dag$,\ Zhangdui Zhong$^\dag$}
\IEEEauthorblockA{$^*$Institute of Network Coding, CUHK\\$^\dag$State Key Laboratory of Rail Traffic Control and Safety, BJTU\\Email: fgwang@inc.cuhk.edu.hk,\ rtxu@bjtu.edu.cn,\ zhdzhong@bjtu.edu.cn}}
\maketitle

\begin{abstract} \label{abs}
Kolmogorov-Smirnov (K-S) test--a non-parametric method to measure
the goodness of fit, is applied for automatic modulation
classification (AMC) in this paper. The basic procedure involves
computing the empirical cumulative distribution function (ECDF) of
some decision statistic derived from the received signal, and
comparing it with the CDFs of the signal under each candidate
modulation format. The K-S-based modulation classifier is first
developed for AWGN channel, then it is applied to OFDM-SDMA systems
to cancel multiuser interference. Regarding the complexity issue of K-S modulation
classification, we propose a low-complexity method based on the
robustness of the K-S classifier.  Extensive simulation results
demonstrate that compared with the traditional cumulant-based
classifiers, the proposed K-S classifier offers superior
classification performance and requires less number of signal
samples (thus is fast).
\end{abstract}


%
\IEEEpeerreviewmaketitle

\section{Introduction}  \label{sec_int}
Automatic modulation classification is a procedure performed at the
receiver based on the received signal before demodulation when the
modulation format is not known to the receiver. It plays a key role
in various tactical communication applications. It also finds
applications in emerging wireless communication systems that employ
interference cancellation techniques -- in order to demodulate and
cancel the unknown interfering user's signal, its modulation format
needs to be classified first.

The feature-based modulation classification methods are popular, and
they base on feature extraction and decision
\cite{moment92}-\cite{2ndstat02}. The most widely used feature is
the cumulant. It can be used to classify many different modulation
types by high-order statistic cumulants \cite{hos00}. It is simple
to implement and can achieve nearly optimal performance with large
number of samples \cite{novel08}. For example, the fourth-order
cumulant can be used to classify various low-order modulations. For
classifying higher-order constellations, a higher-order cumulant is
needed. An accurate estimate of the higher-order cumulant of the
signal requires a large number of signal samples. Most of the
existing works on modulation classification focus on the additive
white Gaussian noise (AWGN) channel. A few works have considered
fading and multipath channels \cite{novel08}, \cite{freqsel00}.
However, effective modulation classifier with less signal samples
remains a challenge.

In this paper, we propose to employ the Kolmogorov-Smirnov (K-S)
test \cite{ks51} for modulation classification. The K-S test is a
non-parametric statistical method to measure the goodness of fit.
From the received signal, we compute the empirical cumulative
distribution function (CDF) of certain decision statistic. A priori
we also compute the CDF of the same decision statistic under each
candidate modulation format. The modulation format that results in
the minimum of the maximum distance between its CDF and the observed
empirical CDF is the final decision. We develop K-S classifiers
based on quadrature amplitude decision statistics, then apply it to
OFDM-SDMA systems to cancel the multiuser interference. Regarding
high complexity involved by CDF calculation, we propose a low-complexity
method based on the robustness of K-S classifier\footnote{It is worth to point out that during the preparation of this paper's presentation we discover the impressive work in \cite{DBLP:journals/corr/abs-1012-5327} that was submitted recently and investigates low complexity issue of modulation classification.}. We provide
extensive simulation results to demonstrate the performance gain of
the proposed K-S classifiers over the cumulant-based classifiers.
The remainder of this paper is organized as follows. In Section II
we provide some background on modulation classification and on the
K-S test. In Section III, we develop the K-S-based modulation
classifiers based on K-S test, and its corresponding low-complexity method.
Section IV is devoted to an
application of K-S classifier in OFDM-SDMA system for interference
signal recognition and cancellation. Simulation results are provided
in Section V. Section VI concludes the paper.

\section{Background}  \label{sec_sys}
\subsection{Automatic Modulation Classification}
Consider the following discrete-time additive white noise channel
model
\begin{equation}
\label{model} y_n = x_n + w_n, \qquad n = 1,\cdots,N,
\end{equation}
where $x_n$, $y_n$ and $w_n$ are respectively the complex-valued
transmitted modulation symbol, the received signal, and the noise
sample at time $n$. The transmitted symbols $\{x_1,\cdots,x_N\}$ are
drawn from an unknown constellation set $\mathcal{M}$ which in turn
belongs to a set of possible modulation formats
$\{\mathcal{M}_1,\cdots,\mathcal{M}_K\}$. The modulation
classification problem refers to the determination of the
constellation set $\mathcal{M}$ to which the transmitted symbols
belong based on the received signals $\{y_1,\cdots,y_N\}$. There are
two major approaches in the literature to solving the above
modulation classification problem. In the likelihood-based methods
\cite{survey07}, \cite{alrt00}, some form of the likelihood for each
modulation format is calculated by making certain assumption on the
data sequence. The classification decision then corresponds to the
modulation with the largest likelihood value. These methods are
typically computationally very expensive. Moreover, they require the
knowledge of the various channel parameters and become ineffective
in the presence of unknown channel impairment such as fading, phase
and frequency offsets, and non-Gaussian interference/noise.

The more popular and low-complexity approach to automatic modulation
classification is based on cumulant \cite{hos00}, \cite{novel08}.
Specifically, for the system given by (\ref{model}), we calculate
the normalized sample fourth-order cumulant of the received signal
$\{y_n\}$ as
\begin{equation}
\hat C =
\frac{\mathbb{E}\{|y|^4\}-|\mathbb{E}(y^2)|^2-2\mathbb{E}^2\{|y|^2\}}{\{\mathbb{E}\{|y|^2\}-\sigma^2\}^2}.
\end{equation}
The modulation whose theoretical cumulant \cite{survey07} is closest
to $\hat C$ is then the classification decision. The fourth-order
cumulants can be used to classify 4-QAM, 16-QAM and 64-QAM
modulations. For even higher-order modulations, the difference
between the cumulants becomes small, which leads to low
classification accuracy. A higher-order cumulant should be used to
classify these modulations, with a considerably increased
computational complexity.

\subsection{Kolmogorov-Smirnov (K-S) Test}
The Kolmogorov-Smirnov (K-S) test is a non-parametric test of
goodness of fit for the continuous cumulative distribution of the
data samples \cite{ks51}, \cite{nonpara_stat}, \cite{nrc92}. It can
be used to approve the null hypothesis that two data populations are
drawn from the same distribution to a certain required level of
significance. On the other hand, failing to approve the null
hypothesis shows that they are from different distributions.

In this paper we consider the one sample K-S test. In the test, we
are given a sequence of i.i.d. real-valued data samples $z_1,
z_2,\cdots,z_N$ with the underlying cumulative distribution function
(CDF) $F_1(z)$, and a hypothesized distribution with the CDF
$F_0(z)$. The null hypothesis to be tested is \be H_0:\quad F_1 =
F_0 \ee

The K-S test first forms the empirical CDF from the data samples
\begin{equation}
\label{1d_ecdf} \hat F_1(z)\triangleq \frac{1}{N}
\sum\limits_{n=1}^{N} {\mathbb{I}(z_n \leq z)},
\end{equation}
where $\mathbb{I}(\cdot)$ is the indicator function, which equals to
one if the input is true, equals to zero otherwise. The largest
absolute difference between the two CDF's is used as the
goodness-of-fit statistic, given by
\begin{equation}
D \triangleq \sup \limits_{z\in \mathbb{R}}|F_1(z)-F_0(z)|,
\end{equation}
and in practical, it is calculated by
\begin{equation}
\label{stat_1d} \hat D = \max \limits_{1 \leq n \leq N}|\hat
F_1(z_n)- F_0(z_n)|.
\end{equation}
The significance level $\hat \alpha$ of the observed value $\hat D$
is given by
\begin{align}
\label{alphaCV} \hat \alpha  \triangleq P(D>\hat
D) &=Q\left([\sqrt{N}+0.12+\frac{0.11}{\sqrt{N}}]\hat D\right),\\
\text{with}\ Q(x) &\triangleq 2\sum
\limits_{m=1}^{\infty}(-1)^{m-1}e^{-2m^2x^2}.
\end{align}
The hypothesis $H_0$ is rejected at a significance level $\alpha$ if
$\hat \alpha=P(D>\hat D)<\alpha$.

\section{K-S-based Modulation Classification}  \label{sec_ks}
Consider the signal model in (\ref{model}). In this section, we
assume that the i.i.d. noise samples follow the complex Gaussian
distribution, i.e., $w_n \sim \mathcal{N}_c(0,\sigma^2)$; that is,
the real and imaginary components of $w_n$ are independent and have
the same Gaussian distribution
$\mathcal{N}_c(0,\frac{\sigma^2}{2})$. To classify the modulation
based on the received signals $\{y_n\}$ using the K-S test, we first
form a sequence of decision statistics $\{z_n\}$ from $\{y_n\}$,
where $z_n$ can be either the magnitude, or the phase, or the real
and imaginary components of $y_n$, and then compute the
corresponding empirical CDF $\hat F_1$. In the meantime, for each
possible modulation candidate $\mathcal{M}_k$, we can obtain either
the CDF $F^k_0$ for $\{z_n\}$. The K-S statistic is then calculated
by
\begin{equation}
\label{9} \hat D = \max \limits_{1 \leq n \leq N}|\hat F_1(z_n)-
F_0^k(z_n)|,\quad k=1,2,\cdots,K.
\end{equation}
The decision on the modulation is given by the minimum K-S
statistic, i.e.,
\begin{equation}
\label{10} \hat k=\arg \min \limits_{1 \leq k \leq K} \hat D_k.
\end{equation}
Moreover, recall that associated with each K-S statistic $\hat D_k$,
there is a significance level $\hat \alpha_k \triangleq P(D > \hat
D_k | \mathcal{M}_k)$, computed by (\ref{alphaCV}). The normalized
$\{\hat \alpha_k\}$ can be used to give a ``soft" decision on the
modulation, that is, the probability that the modulation
$\mathcal{M}_k$ is used approximately $q_k \triangleq \hat
\alpha_k/\sum_{\imath=1}^{I}\hat \alpha_\imath,\ k = 1,\cdots,K$. In
what follows, we discuss the decision statistics $\{z_n\}$ for
different modulation formats, and the corresponding CDF $F_0$.

We consider the quadrature amplitude modulation (QAM) formats, e.g.,
4-QAM, 16-QAM, and 64-QAM. The set of signal points of unit-energy
constellations for these modulations are given by
$\mathcal{M}_{4-\text{QAM}}=\{\frac{1}{\sqrt{2}}(a+b\jmath)|a,b=-1,1\}$,
$\mathcal{M}_{16-\text{QAM}}=\{\frac{1}{\sqrt{10}}(a+b\jmath)|a,b=-3,
-1, 1, 3\}$,
$\mathcal{M}_{64-\text{QAM}}=\{\frac{1}{\sqrt{42}}(a+b\jmath)|a,b=-7,
-5, -3, -1, 1, 3, 5, 7\}$, where $\jmath=\sqrt{-1}$.

We suggest a quadrature-based K-S classifier, which is first proposed in our work of \cite{ks10}. Since for QAM input
signals, the real and imaginary components of the received signal
$y_n$ are independent and have identical distributions, we can also
use them directly as the decision statistics. That is, from the $N$
received signals samples $y_1,\cdots,y_N$, we form a sequence of
$2N$ samples of decision statistic
\[
z_{2n-1}=\mathfrak{R}\{y_n\},\quad z_{2n}=\mathfrak{I}\{y_n\},\ n =
1,\cdots,N.
\]
Then we have $z_n \stackrel{\text{i.i.d.}}{\sim}
\mathcal{N}(0,\frac{\sigma^2}{2})$. Hence the CDF under modulation
$\mathcal{M}_k$ is given by
\begin{equation}
F_{0}^{k}(z)=1-\frac{1}{\sqrt{|\mathcal{M}_k|}}\sum \limits_{x\in
\mathfrak{R}\{\mathcal{M}_k\}}Q_0\left(\frac{\sqrt{2}(z-x)}{\sigma}\right),\
z\in\mathbb{R},
\end{equation}
where $Q_0(a)$ is Gaussian Q-function, and
$\mathfrak{R}\{\mathcal{M}_k\}$ denotes the set of real components
of the signal points in $\mathcal{M}_k$. The K-S test in
(\ref{9})-(\ref{10}) can be performed using $\hat F_1$ and
$\{F_0^k\}$ on the samples $z_1,\cdots,z_{2N}$.

Due to the complicated CDF expressions, it is computationally expensive to calculate CDF at each samples. Since the K-S classifier has the property of robustness, which will be proved in Fig. 1, i.e. the correct classification performance is robust to SNR mismatch. With respect to the robustness, we can quantize the received SNR with different granularity. In a certain quantization, one CDF curve is stored for each SNR, where the curves are also comprising of discrete points which are denser than the SNR interval. All the curves can be computed offline. With the storage of these CDF curves, we can avoid the complicated computation by looking up tables. Thus, all involved computations are only ECDF calculation and comparison operation.

\begin{figure}[!t]
\centering
\includegraphics[width=3.5in]{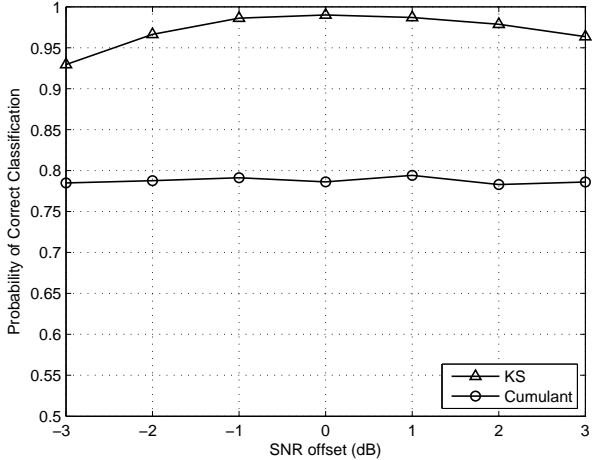}
\caption{QAM modulation (4-QAM, 16-QAM, 64-QAM) classification
performance versus SNR offset. SNR= 15dB.} \label{review2}
\end{figure}

\section{Application: Interference Cancellation in OFDM-SDMA Systems}
We next consider an application of modulation classification in the
context of interference cancellation in an OFDM system employing
multiple receive antennas. Specifically, assuming the OFDM receiver
is equipped with two receive antennas, which makes it possible to
have two users simultaneously transmitting data -- the so-called
space-division multiple-access (SDMA) \cite{Boudreau09}. That is,
the received signal at the $\ell$-subcarrier of the $n$-th OFDM word
is given by
\begin{equation}
\label{12}
\begin{array}{l}
\bY_\ell(n)= \bH_\ell X_\ell(n)+\bG_\ell X'_\ell(n)+\bW_\ell(n),\\
\qquad \ \ell=1,\cdots,P;\ n=1,\cdots,N,
\end{array}
\end{equation}
where $\bY\in \mathbb{C}^{2\times 1}$ denotes the received signals at
the two receive antennas; $\bH\in \mathbb{C}^{2\times 1}$ denotes
the channels between the desired user's transmitter and the receive
antennas; $\bG\in \mathbb{C}^{2\times 1}$ denotes the channels
between the interfering user's transmitter and the receive antennas;
$\bW \sim \mathcal{N}_c(\mathbf{0},\sigma^2 \bI)$ is the received
noise sample vector.

\begin{figure}[!t]
\centering
\includegraphics[width=3.5in]{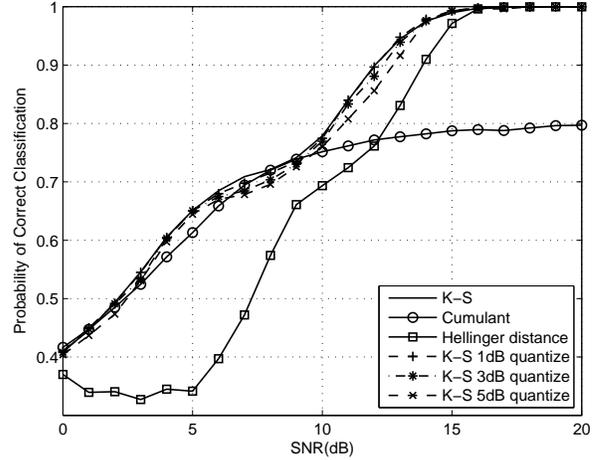}
\caption{QAM modulation \{4-QAM, 16-QAM, 64-QAM\} classification
performance in AWGN channels. The number of samples N = 100.}
\label{awgnqam100}
\end{figure}

The receiver aims to demodulate the desired user's
symbols $\{X_\ell(n)\}$. It is assumed that the receiver knows the
modulation of the desired user, but not that of the interferer. It
is also assumed that the receiver can estimate the channels of both
the desired user and the interferer, $\{\bH_\ell,\bG_\ell\}$. A
simple receiver scheme is to apply a linear MMSE filter to the
received signal $\bY_\ell(n)$ to suppress the interferer, and
demodulate $X_\ell(n)$ based on the output of this filter. A more
powerful receiver scheme is to employ interference cancellation.
That is, we first demodulate the interferer's symbols
$\{X'_\ell(n)\}$ and then subtract the interferer's signals from the
received signals. Finally we demodulate the desired user's symbols
based on the post-cancellation signals. In order to estimate the
interferer's symbols, we must first classify the modulation format
used by the interferer. Hence the interference cancellation receiver
consists of the follow steps.

\begin{itemize}
\item  For each subcarrier, apply a linear MMSE filter $\bM_\ell = \alpha_\ell\left(\bH_\ell\bH_\ell^H+\sigma^2\bI\right)^{-1}\bG_\ell$ to the received signal to suppress the desired signal
\[
\gamma'_\ell(n) = \bM^H_\ell \bY =
\alpha_\ell\bG^H_\ell(\bH_\ell\bH_\ell^H+\sigma^2\bI)^{-1}\bY_\ell(n),
\]
\be \text{with}\ \alpha_\ell =
[\bG_\ell^H(\bH_\ell\bH_\ell^H+\sigma^2\bI)^{-1} \bG_\ell
]^{-1}.\qquad \ee By the choice of $\alpha_\ell$ in (37), we can
write \be \label{14} \gamma'_\ell(n)=X'_\ell(n)+w_\ell(n), \ee where
$w_\ell(n)=\bM^H_\ell(\bH_\ell X_\ell(n)+\bW_\ell(n))$ contains the
residual desired user's signal and noise. The distribution of
$w_\ell(n)$ can be accurately modeled as Gaussian with zero mean and
variance $\tilde \sigma^2=|\bM_\ell^H
\bH_\ell|^2+\sigma^2|\bM_\ell|^2$.
\item  Based on the linear MMSE filter output (\ref{14}), which is an AWGN model, we can apply the
K-S classifiers discussed in Section \ref{sec_ks} to classify the
modulation format for the $l$-th subcarrier group, $l =
1,\cdots,\frac{P}{p}$.
\item  Once the modulation format of the interferer on each subcarrier
group is estimated, we can demodulate the interferer's symbols
$\{X'_\ell(n)\}$ based on the linear MMSE filter output (\ref{14}).
\item  Next we perform interference cancellation on each subcarrier,
followed by a matched-filtering for the desired users's signals,
i.e., \be \beta_\ell(n)=\bH^H_\ell \left(\bY_\ell(n)-\bG_\ell \hat
X'_\ell(n) \right) \ee Finally the desired user's symbol $X_\ell(n)$
is demodulated from $\beta_\ell(n)$.
\end{itemize}

\begin{figure}[!t]
\centering
\includegraphics[width=3.5in]{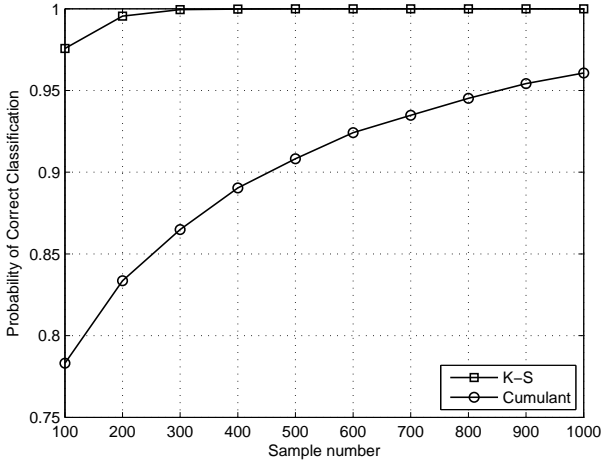}
\caption{QAM modulation \{4-QAM, 16-QAM, 64-QAM\} classification
performance versus sample size in AWGN channels. SNR= 14dB.}
\label{flatqam100}
\end{figure}

\section{Simulation Results}
\label{sec_num}

In this section, we provide simulation results to compare the
performance of the proposed K-S-based modulation classifier with
that of the cumulant-based one. For the QAM
modulations, we will consider the set $\{$4-QAM, 16-QAM, 64-QAM$\}$
in AWGN channel.
The channel model is given by (1) with $w_n \sim
\mathcal{N}_c(0,\sigma^2)$. The signal-to-noise ratio (SNR) is
defined as $1/\sigma^2$. The 4-th order cumulants are used. The number of received signal samples used is $N$ =
100. In Fig. 1, consider the SNR mismatch at the receiver. If the noise power is
half or twice as the original value, there is 3dB or -3dB SNR
mismatch. We compare the robustness of K-S and cumulant classifiers. In
the whole range, K-S classifier is always much better than the cumulant one,
although there is degradation on large value of SNR offset. Cumulant
method is robust in the range, however, its performance is not
satisfactory with such few sample number.
The classification performance of various
classifiers in AWGN channels for QAM modulations is shown in Fig. 2, including quantized K-S classifier. We also show the performance of the
Hellinger-distance-based classifier \cite{Huo98asimple} which has a
very high complexity. It is seen that for such a small sample size,
at high SNR, the cumulant-based methods exhibit a ceiling on the
classification probability around 0.8. However, the K-S-based
classifier monotonically improves the classification performance as
the SNR increases and it significantly outperforms the
cumulant-based classifier at high SNR. The Hellinger-distance-based
classifier performs worse than the cumulant method in the low SNR
region and in the high SNR region it performs worse than the K-S
quadrature method. As is shown in Fig. 2, the performance suffers more as the SNR quantized interval increases, where the quantized interval of each CDF curve is 0.01 and the scale is between -4 to 4, and larger values are ignored since they are with trivial probability. With respect to the performance degradation at -3dB and 3dB SNR mismatch in Fig. 1, K-S classifier is still superior to the other two classifiers even with 5dB quantized interval.
The classification performance for QAM
modulations as a function of the sample size is shown in Fig. 3.

We next consider the effect of modulation classification on the
performance of interference cancellation in an OFDM system employing
multiple receive antennas. The signal model is given by (\ref{12}).
Again the 3GPP channel model is used to generate the multipath
channels for multiple antennas \cite{Channel996}, \cite{996}. There
are 512 subcarriers in one OFDM symbol, which are shared by both the
desired user and the interferer. For simplicity we assume that the
same QAM modulation is employed on all subcarriers for each user.
Hence modulation classification is based on samples from one OFDM
symbol. We assume that the channels of both the desired user and the
interferer are known. In Fig. 4 and Fig. 5 we compare the bit error
rate (BER) performance of four receivers, namely, the linear MMSE
receiver, the interference cancellation receivers using the K-S
modulation classifier and the cumulant-based modulation classifier,
respectively, and an ``ideal" receiver that is assumed capable of
completely removing the interferer's signal (and hence achieving
single-user performance.) Note that the last receiver performance
serves as a lower bound to the performance of any practical
receiver. In Fig. 4 the desired user employs 16-QAM whereas in Fig.
5 the desired user employs 64-QAM. It is seen that performance of
the receiver that uses the cumulant-based classifier is even worse
than that of the linear MMSE receiver. This is because even with a
sample size of 512, the cumulant-based classifier has a relatively
low accuracy in detecting the modulation and with the wrong
modulation information, the interference cancellation receiver
actually enhances the interference. On the other hand, the receiver
that employs the K-S classifier exhibits performance that is close
to the ideal receiver performance, and offer a gain of 2dB and 3dB
respectively at the BER of 0.01 compared to the linear MMSE
receiver.

\section{Conclusion}
We have proposed a new modulation classification technique based on
the Kolmogorov-Smirnov (K-S) test, for classifying different QAM
modulation formats. The basic procedure involves computing the ECDF
of some decision statistic derived from the received signal, and
comparing it with the CDFs of the signal under each candidate
modulation format. Compared with the popular cumulant-based
modulation classifiers, the proposed K-S classifiers offer faster
(i.e., requiring less number of signal samples) and superior
performance. Regarding the complexity issue, we propose the low complexity method based on the robustness of K-S classifier. Moreover, the K-S classifier offers a method of
interference cancellation in OFDM-SDMA systems.

\begin{figure}[!t]
\centering
\includegraphics[width=3.5in]{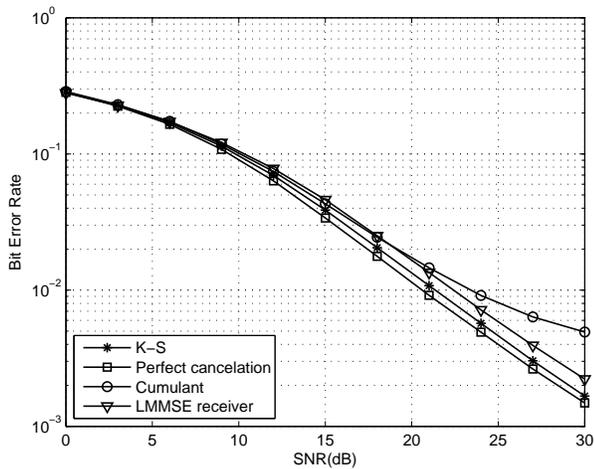}
\caption{BER performance of OFDM-SDMA receivers. The desired user
employs the 16-QAM modulation; and the interferer employs
modulations from \{4-QAM, 16-QAM, 64-QAM\}.} \label{fig8}
\end{figure}

\section*{Acknowledgment}

This work is supported by
Program for Changjiang Scholars and Innovative Research Team in University under Grant No. IRT0949; the Joint Funds of State Key Program of NSFC (Grant No. 60830001);
Program for New Century Excellent Talents in University under Grant
NCET-09-0206; the Key Project of State Key Lab. of Rail
Traffic Control and Safety under Grant RCS2008ZZ006 and RCS2008ZZ007.

\begin{figure}[!t]
\centering
\includegraphics[width=3.5in]{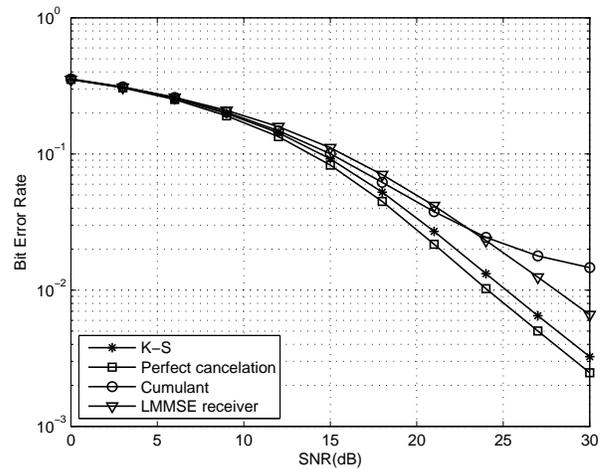}
\caption{BER performance of OFDM-SDMA receivers. The desired user
employs the 64-QAM modulation; and the interferer employs
modulations from \{4-QAM, 16-QAM, 64-QAM\}.} \label{fig9}
\end{figure}

\bibliographystyle{IEEEtran}
\bibliography{ks_amc}

\end{document}

%% file: pream_bm.tex
\newtheorem{prop}{Proposition}

\newtheorem{cor}{Corollary}

\newtheorem{lm}{Lemma}

\newtheorem{thm}{Theorem}

\newcommand{\be}{\begin{eqnarray}}
\newcommand{\ee}{\end{eqnarray}}
\newcommand{\benn}{\begin{eqnarray*}}
\newcommand{\eenn}{\end{eqnarray*}}
\def\IR{\rm I \kern-0.20em R}
\newcommand{\utwi}[1]{\mbox{\boldmath $ #1$}}

\newcommand{\bthm}{\begin{thm}}
\newcommand{\ethm}{\end{thm}}

\newcommand{\bcor}{\begin{cor}}
\newcommand{\ecor}{\end{cor}}
\newcommand{\bprop}{\begin{prop}}
\newcommand{\eprop}{\end{prop}}
\newcommand{\blm}{\begin{lm}}
\newcommand{\elm}{\end{lm}}
\newcommand{\beq}{\begin{equation}}
\newcommand{\eeq}{\end{equation}}
\newcommand{\ber}{\begin{eqnarray}}
\newcommand{\eer}{\end{eqnarray}}

\newcommand{\bproof}{\begin{proof}}
\newcommand{\eproof}{\end{proof}}

\newcommand{\bit}{\begin{itemize}}
\newcommand{\eit}{\end{itemize}}
\newcommand{\ben}{\begin{enumerate}}
\newcommand{\een}{\end{enumerate}}
\newcommand{\bdesc}{\begin{description}}
\newcommand{\edesc}{\end{description}}
\newcommand{\beqarrn}{\begin{eqnarray*}}
\newcommand{\eeqarrn}{\end{eqnarray*}}
\newcommand{\bproofof}{\begin{proofof}}
\newcommand{\eproofof}{\end{proofof}}
\newenvironment{rem}{\begin{trivlist}\item[]{\bf
Remark:}\hspace{4mm}}{\end{trivlist}}
\newcommand{\brem}{\begin{rem}}
\newcommand{\erem}{\end{rem}}
\newenvironment{rems}{\begin{trivlist}\item[]{\bf
Remarks}\begin{itemize}}{\end{itemize}\end{trivlist}}
\newcommand{\brems}{\begin{rems}}
\newcommand{\erems}{\end{rems}}
\newtheorem{fact}{Fact}
\newcommand{\bfact}{\begin{fact}}
\newcommand{\efact}{\end{fact}}
\newtheorem{examp}{Example}
\newcommand{\bexamp}{\begin{examp}\rm}
\newcommand{\eexamp}{\end{examp}}
\newtheorem{defn}{Definition}
\newcommand{\bdefn}{\begin{defn}\rm}
\newcommand{\edefn}{\end{defn}}

\newtheorem{alg}{Algorithm}
\newcommand{\balg}{\begin{alg}}
\newcommand{\ealg}{\end{alg}}

\newtheorem{prob}{Problem}
\newcommand{\bprob}{\begin{prob}}
\newcommand{\eprob}{\end{prob}}

\newcommand{\bvtm}{\begin{verbatim}}
\newcommand{\bfig}{\begin{figure}}
\newcommand{\efig}{\end{figure}}
\newcommand{\bcen}{\begin{center}}
\newcommand{\ecen}{\end{center}}

\long\def\comment#1{}

\def \n2{{N_0 \over 2}}

\def \h5{\hspace{0.5in}}

\newcommand{\bG}{{\utwi{G}}}
\newcommand{\bH}{{\utwi{H}}}
\newcommand{\bI}{{\utwi{I}}}

\newcommand{\bM}{{\utwi{M}}}

\newcommand{\bW}{{\utwi{W}}}

\newcommand{\bY}{{\utwi{Y}}}